\documentclass[preprint,showpacs,preprintnumbers,amsmath,amssymb]{revtex4}

\usepackage{graphicx}
\usepackage{dcolumn}
\usepackage{bm}

\begin{document}
\title{Hiding Under the Carpet: a New Strategy for Cloaking}

\author{Jensen Li, J. B. Pendry}

\affiliation{Blackett Laboratory, Imperial College London, London SW7 2AZ}

\begin{abstract}
A new type of cloak is discussed: one that gives all cloaked objects the 
appearance of a flat conducting sheet. It has the advantage that none of the 
parameters of the cloak is singular and can in fact be made isotropic. 
It makes broadband cloaking in the optical frequencies one step closer.
\end{abstract}

\maketitle

Transformation optics \cite{Ward:1996,Schurig:OptExp2006,Pendry:2006,Schurig:2006} can be used to design a cloak 
of invisibility. In essence the cloak makes its contents appear to be very 
small and hence invisible. However there are three distinct topological 
possibilities: the cloaked object can be crushed to a point, to a line, or 
to a sheet. In the process of crushing the object becomes infinitely 
conducting but this does not present a problem for the first case because 
scattering from even highly conducting small objects vanishes with the size 
of object. The same is true of a very thin object, i.e. a wire, because very 
thin wires have very large inductance and are therefore invisible to 
radiation. This is not true of the third possibility. Obviously a conducting 
sheet is highly visible unless of course it sits on another conducting sheet 
(the `carpet' of the title). Although this third possibility has limited 
cloaking potential it does have the considerable advantage that the 
parameters of the cloak need not be singular and, as we shall show, can be 
isotropic.

Metamaterials are usually used as the building blocks for transformation optics as they can attain a wide range of permittivity and permeability. 
Several works investigated the cylindrical cloak in full wave analysis and the first experiment in verifying the cloak is done in microwave. \cite{Cummer:2006,Zolla:2007,Ruan:2007,Zhang:2007}
There are other approaches targeting the subwavelength regime \cite{Alu:2005,Milton:2006} 
and the geometrical optics regime \cite{Leonhardt:2006} for cloaking as well. 
Here, we concentrate on the transformation optics approach which is driven by coordinate transformation 
and has a large variety of applications.\cite{Rahm:1,Chen:2007}

There is still a general quest to push the working frequency to visible regime for more applications. 
However, scaling down the magnetic resonating structures for the metamaterials to optical frequencies causes severe absorption. \cite{Stefan:2004,Shalaev:2005,Zhang:2005,Dolling:2006,Dolling:2007}
One way to avoid it is to use the reduced material parameters in the cloak so that only anisotropic 
metamaterials with electric resonating elements are needed. \cite{Cai:2007} However, as long as we are using 
metamaterials with resonating structures, materials absorption still cannot be neglected. 

In this work, instead of the complete cloak, we will consider a cloak to mimic a flat ground plane.
We will show that it does not require singular values for the material parameters, i.e. the range for the permittivity and the 
permeability is much smaller than in the case of a complete cloak. Moreover, by choosing a suitable coordinate transform, the 
anisotropy of the cloak can be minimized to a small value also. 
As a result, we can avoid using metamaterials and just use isotropic dielectrics to construct the cloak.
It greatly reduces absorption and also make broadband cloaking one step closer.

\begin{figure}[htbp]
\centerline{\includegraphics[width=5.0in]{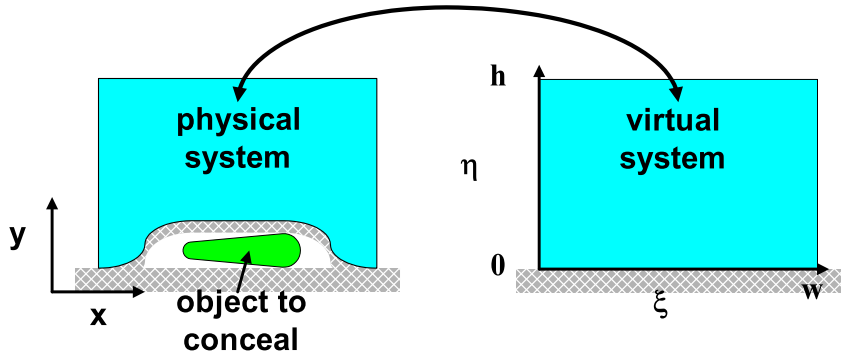}}
\caption{(color online). The general mapping between the virtual system and the physical system. The 
regions in cyan color are transformed into each other. The shaded region at 
the bottom of both domain represents the ground plane (a perfect conductor). The observer perceives the physical system (x-system) as the virtual 
system with a flat ground plane.}
\label{fig1}
\end{figure}

Here, we consider only the 2D wave problem for simplicity, i.e. all the fields are invariant in the 
z-direction. Furthermore, the E-polarization is assumed while the formulation can be easily adopted for the other 
polarization. A ground plane here means a highly reflecting surface made of metal. It is 
regarded as a perfect conductor.
Suppose an object lies on it, we want to design a cloak covering on the object so that 
the observer perceives the system as a flat ground plane again without 
introducing any additional scattering from the object. The object is now 
concealed between the cloak and the original ground plane as shown on the 
left hand side of Fig. \ref{fig1}. We further assume that 
the cloak (shown in cyan color) has a rectangular shape of a width of $w$ and a 
height of $h$ except that the bottom (inner) cloak boundary is curved upwards to leave enough 
space to conceal the object. The whole configuration is termed as the 
physical system with coordinate $\left( {x,y} \right)$ or literally written 
as $\left( {x^1,x^2} \right)$ in indexed notation. The virtual system is the 
configuration the observer perceives. It is shown on the right hand side of 
Fig. \ref{fig1}. Its coordinate is labeled by $\left( {\xi 
,\eta } \right)$ or literally written as $\left( {\xi ^1,\xi ^2} \right)$ in 
indexed notation. In general, we are considering a coordinate transform 
which maps (with sense preserving) a rectangular region ($0 \le \xi \le 
w$,$0 \le \eta \le h)$ in the virtual system to an arbitrary region (the 
cloak in this case) in the physical system. 

To ease our discussion, we introduce the Jacobian matrix $\Lambda $ by

\begin{equation}
\label{eq5}
\Lambda ^i_{i'} = \frac{\partial x^i}{\partial \xi ^{i'}},
\end{equation}

\noindent
and the covariant metric $g$ by

\begin{equation}
\label{eq6}
g_{{i}'{j}'} = \vec {\xi }_{i}' \cdot \vec {\xi }_{j}' \quad \mbox{or}\quad 
g = \Lambda ^T\Lambda ,
\end{equation}

\noindent
where $\vec {\xi }_1 $, $\vec {\xi }_2 $ are the covariant basis vectors of 
the virtual coordinates appearing in the physical system. 
In the application of cloaking, the observer perceives the physical system 
as the virtual system of an isotropic homogeneous medium of permittivity $\varepsilon _{\mbox{ref}}$ and unit permeability.
The corresponding permittivity and permeability in the physical system induced by the coordinate transformation are given by

\begin{equation}
\label{eq10}
\begin{array}{l}
 \varepsilon = \varepsilon_{ref} / \sqrt {\det g} , \\ 
 \left[ {\mu ^{ij}} \right] = \frac{1}{\sqrt {\det g} } \Lambda \Lambda ^T.
 \end{array}
\end{equation}

\noindent
We can write $\mu _T $ and $\mu _L $ be the principal values of the matrix 
representation of the permeability tensor in the physical domain and the 
corresponding refractive indices be $n_T = \sqrt {\mu _L \varepsilon } $ and 
$n_L = \sqrt {\mu _T \varepsilon } $ for the two (local) plane waves 
traveling along the two principal axes. To indicate the extent of 
anisotropy of the physical medium, the anisotropy factor $\alpha $ 
(a function of position) is defined by

\begin{equation}
\label{eq11}
\alpha = \max \left( {n_T / n_L ,n_L / n_T } \right).
\end{equation}

\noindent
By using Eq. (\ref{eq10}) and Eq. (\ref{eq6}), it can be proved that

\begin{equation}
\label{eq12}
\alpha + \frac{1}{\alpha } = \frac{Tr\left( g \right)}{\sqrt {\det g} },
\end{equation}

\noindent
with

\begin{equation}
\label{eq13}
\mu _L \mu _T = 1 .
\end{equation}

\noindent
On the other hand, we can define an averaged refractive index $n$ relative to the reference medium by

\begin{equation}
\label{eq14}
n = \sqrt {n_L n_T } / \sqrt {\varepsilon _{ref} } ,
\end{equation}

\noindent
so that

\begin{equation}
\label{eq15}
n^2 = \frac{\varepsilon }{\varepsilon _{ref} } = \frac{1}{\sqrt {\det g} }.
\end{equation}

Instead of using $\varepsilon $ and $\mu ^{ij}$ to describe the physical medium, 
we now use $\alpha $ and $n$ (related to the ratio and the product of $n_T $ and $n_L )$ which have geometrical meanings in 
terms of the metric. If we imagine there is a very fine rectangular grid in 
the virtual domain with every tiny cell being a square, the mapping (or the coordinate transform) 
transforms this grid to another grid in the physical domain. Every such tiny 
square ($\delta \times \delta )$ in the virtual domain is transformed to a 
parallelogram with two sides $\vec {\xi }_1 \delta $and $\vec {\xi }_2 
\delta $. A smaller anisotropy means a smaller value of $Tr\left( g \right) / \sqrt {\det g} $ 
while a smaller area of the transformed cell ($\sqrt {\det g} \delta 
^2)$ means a larger refractive index $n$.

In cloaking, compression of space in the physical domain essentially makes the cloak anisotropic.
However, our heuristic approach is to minimize the anisotropy induced in the physical medium by 
choosing a suitable coordinate transform. If the anisotropy is small enough, we can simply drop 
this part (by assigning $\alpha = 1$) and only keep the refractive index $n$. In other words, 
the physical medium becomes just a dielectric profile described by Eq. (\ref{eq15}) with unit magnetic 
permeability.

In this work, such an optimal map is generated by minimizing the Modified-Liao functional \cite{Knupp:1994}

\begin{equation}
\label{eq21}
\Phi = \frac{1}{hw}\int_0^w {d\xi } \int_0^h {d\eta } \frac{Tr\left( g 
\right)^2}{\det g},
\end{equation}

\noindent
upon slipping boundary condition. Slipping boundary condition means that each of the four 
bounding edges of the virtual domain must be mapped to the four specified 
boundaries in the physical domain, only up to a sliding freedom. The minimal of this functional occurs 
at the quasiconformal map \cite{Thompson:1998}. 
Without going further for technical proof, we state the result that the quasiconformal map 
actually minimizes not only the average but also the maximum value of $Tr\left( g \right) / \sqrt {\det g} $  in the physical domain, 
i.e. anisotropy minimized.
If we imagine there is a very fine rectangular grid in the virtual domain with 
every tiny cell being a square, the transformed grid in the physical domain 
for the quasiconformal map has the property that every transformed cell is 
a rectangle of a constant aspect ratio $M:m$ where $M$ is called the 
conformal module of the physical domain (a geometrical property of the physical domain once the four boundaries are specified) 
and $m = w / h$ is the conformal module of the virtual domain, i.e.

\begin{equation}
\label{eq17}
\frac{\left| {\vec {\xi }_1 } \right|}{\left| {\vec {\xi }_2 } \right|} = 
\frac{M}{m},\quad \sqrt {\det g} = \left| {\vec {\xi }_1 } \right|\left| 
{\vec {\xi }_2 } \right|.
\end{equation}

\noindent
By substituting Eq. (\ref{eq17}) into Eq. (\ref{eq12}), we have

\begin{equation}
\label{eq18}
\frac{Tr\left( g \right)}{\sqrt {\det g} } = \frac{M}{m} + 
\frac{m}{M},\;\quad \mbox{or}\quad \alpha = \max \left( 
{\frac{M}{m},\frac{m}{M}} \right),
\end{equation}

\noindent
independent of position.

\begin{figure}[htbp]
\centerline{\includegraphics[width=4.0in]{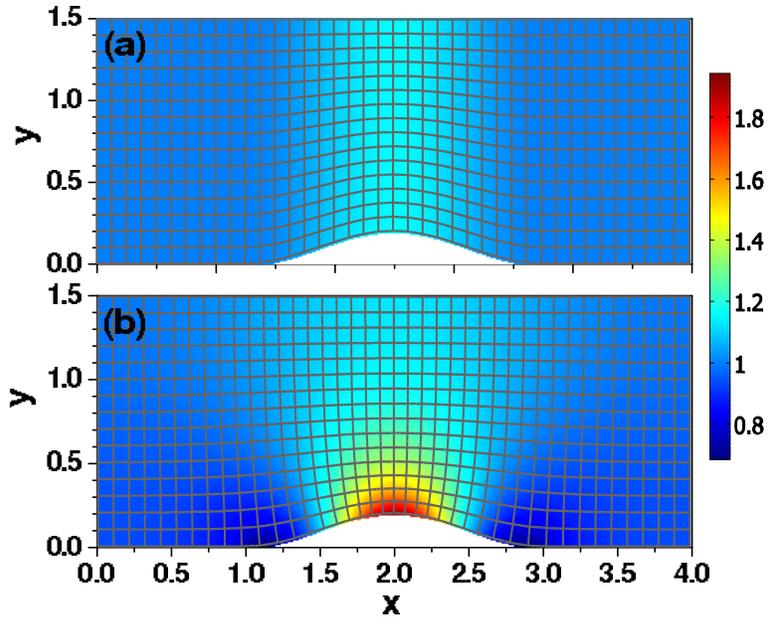}}
\caption{(color online). The transformed grid in physical system with inner cloak boundary specified by Eq. (\ref{eq22}) for 
the (a) transfinite grid and the (b) quasiconformal grid. The color maps show profile $n^2.$}
\label{fig2}
\end{figure}

As a first example, let us map the area of a rectangle bounded by $0 \le \xi ^1 \le 4$, $0 \le \xi ^2 \le 1.5$ in the virtual domain to the same 
rectangle in the physical domain but with the bottom boundary specified by

\begin{equation}
\label{eq22}
y_{\mbox{bottom}} \left( x \right) = \left\{ {{\begin{array}{*{20}c}
 {0.2 cos(\pi x/2)^2} \hfill & {1 \le x \le 3,} \hfill \\
 0 \hfill & {\mbox{otherwise}.} \hfill \\
\end{array} }} \right.
\end{equation}

\noindent
The transformed area in the physical domain is exactly the region of the cloak.
Fig. \ref{fig2}(a) shows the transformed grid in the physical system if a simple 
transfinite interpolation is used to map a regular $40\times 15$ grid in the virtual domain. 
In this case, the grid is just a linear compression in the $y$-direction. The corresponding range of the 
anisotropy factor $\alpha $, obtained from Eq. (\ref{eq12}), ranges from $1$ to 
$1.385$ while $n^2$, obtained from Eq. (\ref{eq15}), 
ranges from $1.0$ to $1.153$. On the other hand, for the generated grid using quasiconformal 
map, shown in Fig. \ref{fig2}(b), the grid lines are orthogonal to each other. The aspect ratio of each cell or the anisotropy factor $\alpha $ becomes  
a constant of $1.042$ while $n^2$ ranges from 0.68 to 1.96. As the area of each cell is 
proportional to the square reciprocal of $n$, the minimum of $n$ occurs on the inner cloak boundary at around $x = 1$ where the cell area is largest 
while the maximum occurs also on the inner cloak boundary at $x = 2$ where the cell area is smallest.
Therefore, we sacrifice $n$ to a larger range at the same time anisotropy is minimized.
Nevertheless, $n$ or $\varepsilon$ is always finite without approaching either zero or infinity. 
It is the result of crushing the object to a conducting plane instead of a line so that there is no singular point in the coordinate transform.
The small anisotropy together with the finite range of $n$ makes realization of the cloak easier.

\begin{figure}[htbp]
\centerline{\includegraphics[width=4.0in]{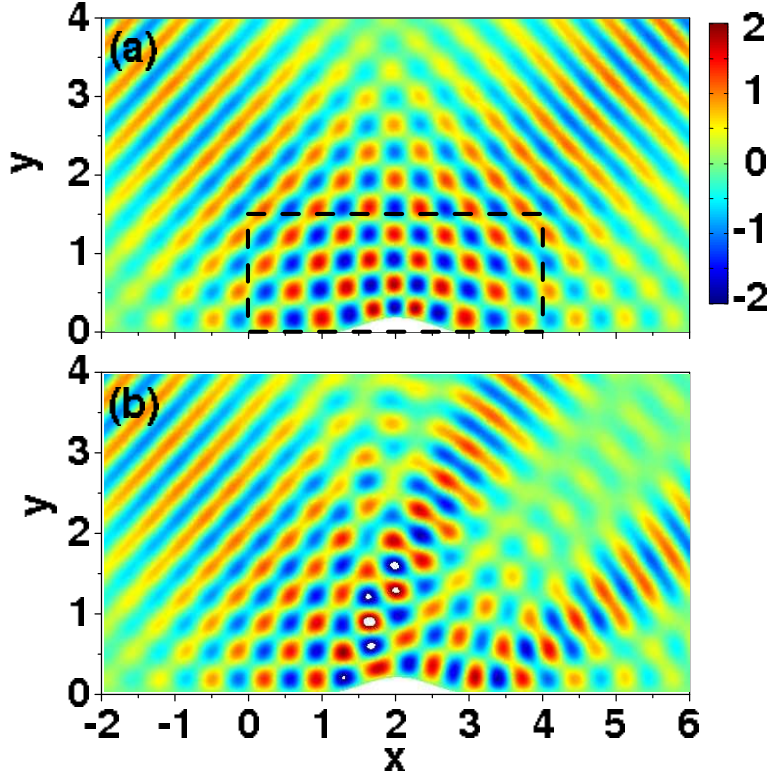}}
\caption{(color online). (a) E-field pattern with the cloak located within the rectangle in dashed line when 
a Gaussian beam is launched at 45 degrees towards the ground plane from the left. (b) E-field pattern when only the object (a reflecting surface of the shape of the inner cloak boundary) is present without the cloak.
All lengths are in ${\mu}m$. The width of the beam is around 4${\mu}m$ at a wavelength of 750nm. The medium above the ground plane and outside the cloak is SiO2 of $n=1.5$}
\label{fig3}
\end{figure}

To test the effectiveness of the designed cloak, suppose the cloak is $4 {\mu}m$  by $ 1.5 {\mu}m$, i.e. the length unit in the forehead discussion is in ${\mu}m$ 
and the $n$ profile given before is defined relative to silica glass ($SiO_2$) of $\varepsilon _{ref}=2.25$.
We ignore the anisotropy in the cloak and only keep the 
part of permittivity. In this case, the permittivity of the cloak varies from around $1.5$ to $4.4$. This range of permittivity can be obtained effectively by etching or drilling 
subwavelength holes of different sizes along the direction of E-field in a high dielectric, e.g. Si. 
Outside the cloak, it is again the silica glass as background material. 
Moreover, the inner surface of the cloak is coated by a highly reflective metal. To the observer, it is perceived as if this is the actual ground plane.
This is relevant to the situation of routing light at our own will in optical integrated circuits.
Suppose a Gaussian beam at a wavelength of 750nm (i.e. 500 nm in $SiO_2$) is launched at an angle of 45 degrees to the object together with the cloak, 
the total E-field pattern (real part) obtained from a FDTD simulation is shown in Fig. \ref{fig3}(a). 
The cloak is within the rectangle in dashed line in the figure.
The field outside the cloak resembles the field as if we only have a flat ground plane. A reflected beam at 45 degrees is clearly seen.
Moreover, the field inside the cloak shows the interference pattern between the incident and the reflected beam. The pattern is simply squeezed upwards if it is compared 
to the field when only a flat ground plane is present.
The corresponding E-field pattern with only the object present is also shown in Fig. \ref{fig3}(b) for comparison. 
Without the cloak, the incident beam is deflected and split into 
two different angles. Therefore, our designed cloak successfully mimics a flat ground plane.

\begin{figure}[htbp]
\centerline{\includegraphics[width=4.0in]{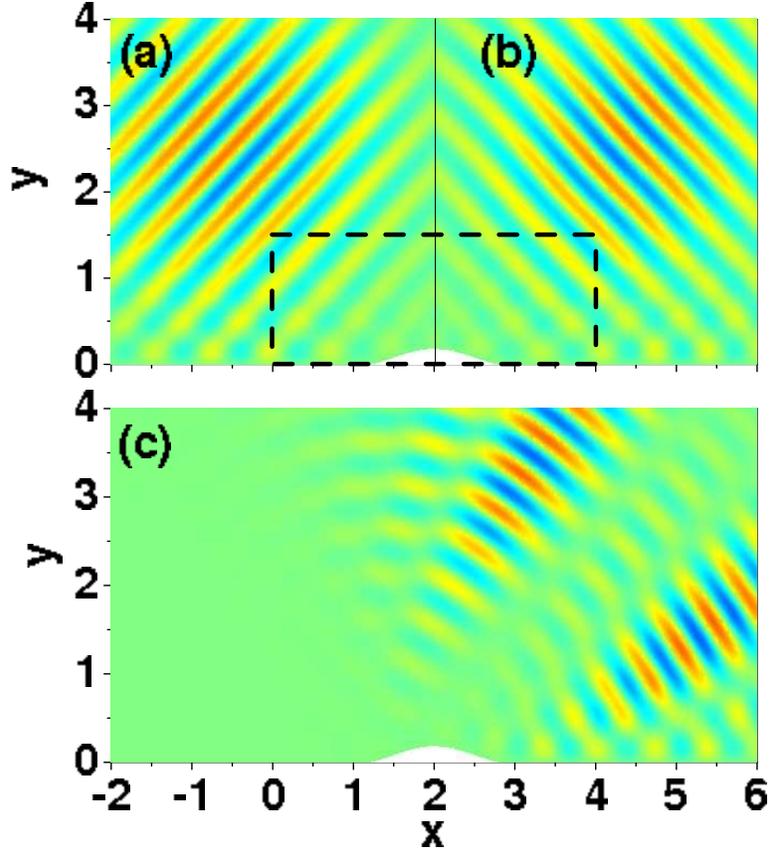}}
\caption{(color online). (a) E-field pattern of the incident wave packet at time $t=0$.
A Gaussian wave packet with $2{\mu}m$ in spatial width and $20fs$ in time width (at a central wavelength of 750nm) is launched at 45 degrees towards the ground plane from the left.
(b) E-field pattern showing the reflected packet from the cloak at time $t=37fs$.
(c) E-field pattern at time $t=37fs$ when the cloak is absent.
All length scales are in ${\mu}m$. The medium above the ground plane and outside the cloak is SiO2 of $n=1.5.$}
\label{fig4}
\end{figure}

We emphasize that our strategy is valid in both geometrical and wave optics. In fact, it is broadband in nature.
Our design works for a wide range of frequencies since the frequency dispersion (and also the losses) of the dielectrics can 
be made very small. 
For example, we can launch a Gaussian wave packet centered at the aforementioned frequency to the object and the cloak.
The transverse width of the packet is set to $2{\mu}m$ with a duration of $20fs$ (8 periods of central frequency). 
It results an incident packet with circular spatial profile at time $t=0$, as shown in Fig. \ref{fig4}(a). 
It is traveling at 45 degrees towards the object. After a duration of 14.8 periods at $t=37fs$, the packet is reflected 
back at 45 degrees and it remains the same size of the Gaussian shape with only a small distortion (or dispersion) as shown in Fig. \ref{fig4}(b).
This is certainly a time-domain cloaking effect. 
For comparison, we have also shown the field pattern at $t=37fs$ if the cloak is absent in Fig, \ref{fig4}(c). The incident wave packet is split into two elongated packets.
There is nearly no distortion for the reflected wave packet in Fig. \ref{fig4}(b). However, if a thinner cloak or bigger object is used, it is expected that 
the distortion will become larger. It is due to a larger anisotropy neglected in our cloak profile. 

In conclusion, a cloak is designed to mimic a flat ground plane. 
The involved coordinate transform induces no singular values in the material profile.
Moreover, the quasiconformal map is the optimal coordinate transform which can minimize the anisotropy of the physical medium. 
The small anisotropy is further neglected so that the cloak can be synthesized by only isotropic dielectrics.
It can relieve the loss issue by avoiding the usage of resonating elements and is a step closer to do broadband transformation optics in optical frequencies. 
The quasiconformal map should also be a convenient technique for both polarizations as well and in other applications in transformation optics.

This work is supported by the Croucher Foundation fellowship from Hong Kong.

\end{document}